\documentclass[pdflatex,sn-nature]{sn-jnl}

\usepackage[utf8]{inputenc}
\usepackage{graphicx}
\usepackage{amsmath,amssymb,amsfonts}
\usepackage{booktabs}
\usepackage{textcomp}

\begin{document}

\title[Source-conditioned polychromatic PSFs]{Broad-band host measurements of little red dots and similar compact nuclei require source-conditioned polychromatic PSFs}

\author*[1]{\fnm{Sergio} \sur{Bonaque-González}}\email{sbonaque@ull.edu.es}

\affil*[1]{\orgdiv{Instituto de Estudios Avanzados IUDEA, Departamento de Física}, \orgname{Universidad de La Laguna}, \orgaddress{\city{San Cristóbal de La Laguna}, \postcode{38206}, \state{Santa Cruz de Tenerife}, \country{Spain}}}

\maketitle
Characterizing compact sources in the early Universe is difficult because light from physically distinct components can overlap within the same diffraction-limited image. A common method uses field stars to model how a single point of light appears and from this estimates the intrinsic structure and contributions of the physical components of the distant object under study. Many important properties inferred for high-redshift objects depend on the accuracy of this method. Using the little red dots (LRDs) analysed by Zhang et al.\cite{zhang2026} as a test case, this work shows that applying the method through broad filters to objects whose spectra differ from those of the field stars can systematically misassign light, overestimating an existing extended component or creating an apparent one where none exists. At the measured level, this error is large enough to alter the inferred sizes, luminosities and masses, and therefore the physical interpretation of these systems. The physical and mathematical origin of the error is established and a correction applicable to other compact distant sources is demonstrated.

Zhang et al. searched for host galaxies around the bright central sources of 217 LRDs\cite{zhang2026}. In outline, they built empirical point-spread functions (PSFs) from field stars and fitted each LRD with the stellar PSF for the unresolved central source together with a Sérsic profile for the possible host galaxy. They subtracted the fitted central source and neighbouring objects, aligned and averaged the residuals to increase the signal-to-noise ratio, and fitted the average with a Sérsic profile. Following a series of controls, the Sérsic component was interpreted as host-galaxy light. It contains 39.6\%, 33.9\%, 23.2\% and 11.8\% of the averaged total flux in four filters of the James Webb Space Telescope Near Infrared Camera (JWST/NIRCam), from F115W to F444W. F444W samples the rest-frame optical at the mean redshift of 6.45 and yields an inferred host half-light radius of approximately 210 parsecs and an apparent magnitude of 27.71 AB. Together with the shorter-wavelength measurements, it gives a stellar mass of $10^{9.02}\,M_\odot$.

The accuracy with which a stellar PSF represents a target depends on how similarly their spectra weight the filter. The polychromatic point-spread function (P-PSF) is the photon-weighted sum of the monochromatic diffraction patterns transmitted by the filter, with their weights determined jointly by the instrument throughput and the source spectrum $f_\nu$. That dependence is strong enough that P-PSF geometry alone separates stellar temperatures differing by 100 to 150~K at a signal-to-noise ratio of order one thousand\cite{bonaque2026}. Importantly, in F444W, field stars are approaching the Rayleigh--Jeans regime, where their normalized spectra vary only weakly with temperature, and consequently form a narrow and stable P-PSF calibration family. However, an LRD at redshift 6.45 lies far outside this stellar family because F444W observes light emitted near 0.60~$\mu$m, where the optical continuum and strong emission lines determine the weighting across the filter. This weighting is summarized by the local spectral slope
\begin{equation}
\beta = \frac{\mathrm{d}\ln f_\nu}{\mathrm{d}\ln \lambda},
\label{eq:spectral_slope}
\end{equation}
which records how strongly a source weights the red end of the wavelength ($\lambda$) range relative to the blue end. Calculated here using blackbody spectra and the median nuclear photometry published by Zhang et al., the sequence from 2,500 to 30,000~K spans only 0.73 in $\beta$, whereas the representative LRD lies 5.66 from a 4,000~K star, almost eight times the entire stellar range.

The ambiguity is independent of the instrument. Consider two nested linear models. The reduced model fits a normalized primary template $v$ and nuisance terms, while the full model adds a normalized second pattern $g$ with non-negative amplitude $b$. Let $\Pi_0$ denote the noise-weighted projection onto the complete reduced model and define $q=g-\Pi_0g$, with $\langle\cdot,\cdot\rangle$ denoting the corresponding pixel inner product. For a noise-free source with normalized primary pattern $u$ and total brightness $\theta$, the fitted amplitude is
\begin{equation}
b = \theta \max\!\left(0, \frac{\langle u, q\rangle}{\langle q, q\rangle}\right).
\label{eq:false_allocation}
\end{equation}
If $u$ belongs to the reduced model, its projection onto $q$ vanishes and $b=0$. If $u$ lies outside the model, any positive projection onto $q$ is assigned to $g$ even when the image contains only the primary source. For fixed pixel weights, the false fraction $b/\theta$ is independent of source brightness. Stacking $N$ objects reduces random noise as $N^{-1/2}$ while preserving any projection shared by the sample. The fit therefore establishes a component along $q$, while its physical origin remains an additional assumption.

In the decomposition used by Zhang et al., $v$ is the empirical stellar P-PSF. Each $g$ is an allowed Sérsic profile convolved with $v$, and $q$ is the part left after fitting the point-source amplitude, position and background. Interpreting a positive $b$ as a host requires the true nuclear P-PSF $u$ of an unresolved LRD to belong to the reduced model. Their matched simulations impose $u=v$, while their stellar controls explore only the narrow family of stellar P-PSFs. Their compact-galaxy control combines intrinsic source structure with any chromatic projection onto $q$ and therefore cannot isolate the missing case. That case is an unresolved source with an LRD spectrum and no host galaxy.

\begin{figure}[htbp]
\centering
\includegraphics[width=\textwidth]{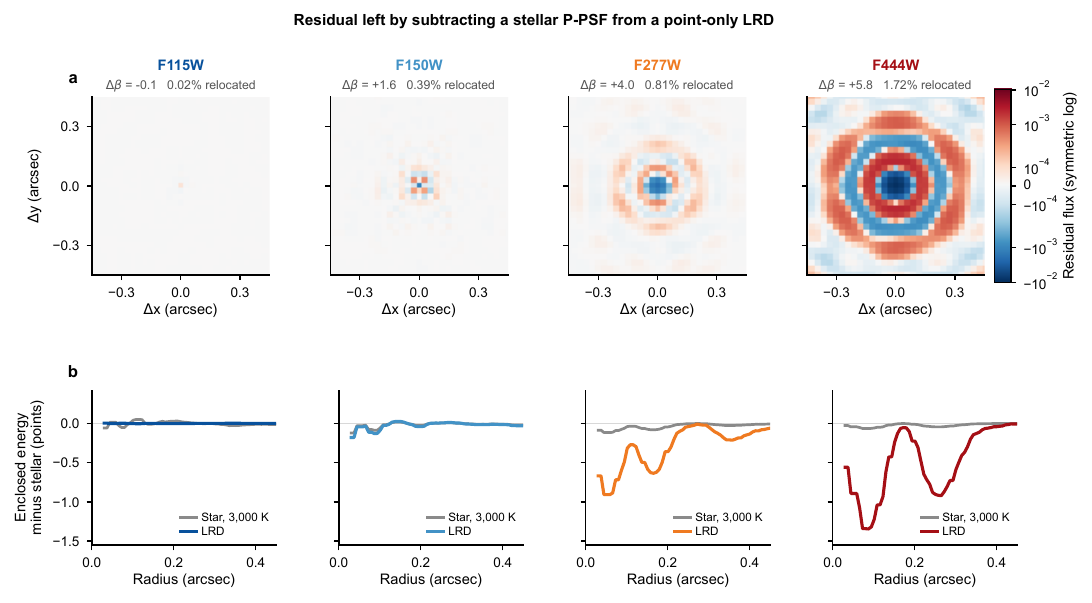}
\caption{\textbf{A point-only LRD produces the wavelength-dependent residual assigned to a Sérsic component.} \textbf{a}, Residual obtained in each filter after subtracting the stellar P-PSF from a source-conditioned LRD P-PSF. Every input contains one unresolved point and zero host flux. Colours use a common symmetric logarithmic scale. Above each panel, $\Delta\beta$ is the in-band slope of the LRD minus that of the stellar mixture from which the reference P-PSF is built, and the percentage is one half of the summed absolute difference between the normalized P-PSFs. \textbf{b}, Enclosed energy relative to the mean stellar reference, in percentage points, for the LRD in colour and for a single field star in grey. Both curves carry the same quantity for different sources. The star shown is the member of the reference mixture that departs from the mean by the most, which is 3,000~K in every filter. Panel colours run from blue to red with increasing wavelength.}\label{fig1}
\end{figure}

To test whether this mismatch can create apparent host light, the decomposition was applied to a synthetic unresolved LRD known to contain no galaxy, as detailed in Methods. The fit nevertheless assigned 0.00\%, 2.62\%, 6.58\% and 18.73\% of its flux to a Sérsic component from F115W to F444W, the last value being 1.59 times their complete 11.8\% component, as shown in Fig.~\ref{fig1} and Table~\ref{tab1}. The underlying P-PSF differences redistribute only 0.02\%, 0.39\%, 0.81\% and 1.72\% of the point-source light, showing that the exchange of flux between the fitted nucleus and the compact Sérsic profile amplifies a small change in the nuclear light distribution into a much larger apparent host fraction. The results remain compatible with genuine hosts because the mismatch produces no false Sérsic allocation in F115W and accounts for only 8\% and 28\% of the reported F150W and F277W components. In F444W, however, it can account for the complete component from which the host size, rest-frame optical luminosity and stellar mass are inferred. Those physical properties therefore require a source-conditioned refit of the original images. 

\begin{figure}[htbp]
\centering
\renewcommand{\figurename}{Extended Data Fig.}
\renewcommand{\thefigure}{1}
\includegraphics[width=\textwidth]{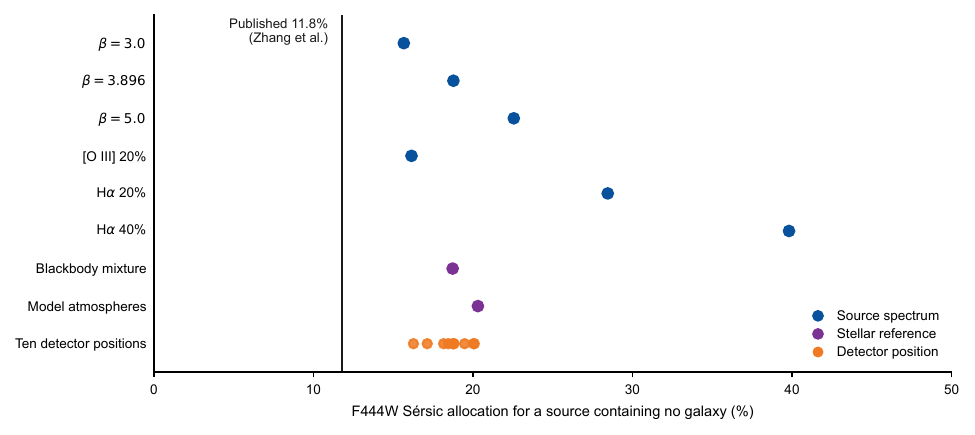}
\caption{\textbf{The F444W allocation for a source containing no galaxy survives every tested change of input.} Each point is one minimum-$\chi^2$ Sérsic allocation obtained by repeating the zero-host construction with a single physical input changed. Blue, six alternative LRD nuclear spectra comprising smooth continua with $\beta$ of 3.0, 3.896 and 5.0, one [O\,\textsc{iii}] case and two H$\alpha$ cases. Purple, the two stellar references, a blackbody mixture spanning 3,000 to 10,000~K and a mixture of Castelli and Kurucz ATLAS9 model atmospheres spanning 4,000 to 8,000~K. Orange, ten detector positions comprising the centre and four off-axis positions on each of NRCA5 and NRCB5. The vertical line marks the 11.8\% F444W component published by Zhang et al. Every tested configuration lies to its right.}\label{edfig1}
\end{figure}

Dedicated controls tested blank sky, realistic backgrounds and noise, alternative LRD spectra, stellar references, detector positions, wavelength sampling, residual stacking and the original galight implementation. These tests identify the chromatic mismatch as the dominant systematic contribution under the conditions examined. Full details are provided in Methods and Extended Data Fig.~\ref{edfig1}.

The correction follows from the same condition. Where the nuclear spectrum is known, the nuclear P-PSF is computed for that spectrum, which places the true pattern inside the model by construction. Where it is not, two diffraction modes derived from a library of LRD P-PSFs let the fitted nucleus adapt without the spectrum of any individual source being supplied. That second form was tested in the two directions that matter. On 3,500 simulated images holding a point source and no galaxy, the conventional model reports a galaxy in $86.17\% \pm 0.58\%$ of them and the corrected one in $4.34\% \pm 0.34\%$. On simulated images holding a genuine galaxy at 5\%, 12\% and 25\% of the total light, the conventional model recovers 18.76\%, 25.10\% and 35.74\% and the corrected one 5.29\%, 12.15\% and 24.68\%. Across the seven spectra and the three host brightnesses tested, the correction neither creates extended emission that is absent nor removes extended emission that is present, and on real COSMOS-Web backgrounds the paired chromatic excess it leaves is consistent with zero. Methods gives the full account, and Extended Data Table~\ref{tabED1} reports the matching test for a template built from a known spectrum.

The catalogue published by Zhang et al. permits a provisional estimate of the astrophysical impact. Correcting each object according to its F277W-to-F444W nuclear slope leaves an aggregate F444W Sérsic fraction of 4.6\%, with a 95\% interval from 3.2\% to 6.3\%. This estimate attributes 61\% of the published 11.8\% component to the P-PSF mismatch and makes the inferred host 2.5 times fainter, corresponding to a dimming of 1.0 magnitude in the rest-frame optical. At fixed stellar mass-to-light ratio, the published mass falls from $10^{9.02}$ to approximately $10^{8.61}\,M_\odot$. Exact values for the flux, size and mass require refitting the original images with source-conditioned P-PSFs.

The correction also makes the host spectrum bluer, because it removes a share that grows with wavelength and nothing at all in F115W, so a refitted population would generally favour a younger or less obscured population and could lower the mass further. Independent spectroscopy points the same way. Sun et al. obtain $10^{8.3}\,M_\odot$ from their median stack of 98 LRDs\cite{sun2026}, against the imaging value of $10^{9.02}$, and the chromatic correction alone moves imaging to $10^{8.61}$. A bias that inflates host light in imaging while leaving spectroscopy untouched predicts that ordering, so the two techniques may be disagreeing for a measurable reason. LRD hosts would then hold substantially less assembled stellar mass than the imaging decomposition reports, and sit elsewhere on the early-galaxy size--mass relation.

These unresolved red points may combine a compact stellar system, an accreting black hole and ionized gas, with recent spectroscopy indicating that reprocessed gas can dominate their rest-frame optical light\cite{sun2026,rusakov2026}. A stellar PSF can divide that light between nucleus and stars incorrectly even when calibrated perfectly across the detector, because the target and the calibration stars weight a broad filter differently. Zhuang and Shen established that such a mismatch overestimates host fluxes, makes the recovered host more compact, and produces a systematic larger than the formal fitting uncertainty\cite{zhuang2024} and the same stellar-template construction underlies other searches for hosts around these objects\cite{chen2025,whalen2026}. Any host detection, size, luminosity or mass obtained by subtracting a stellar PSF from a compact distant source in a broad filter is therefore conditional on the target spectrum, and the published tests of that condition examine the residual itself and not the share of it that the allowed host family absorbs. Source-conditioned P-PSFs make the condition measurable and correctable.

\section{Methods}\label{sec:methods}

\subsection{Nucleus--host decomposition with an empirical stellar PSF}\label{sec:nucleus-host-decomposition}

A common way to separate an unresolved nucleus from its surrounding galaxy is to fit a point source and a galaxy profile simultaneously, using a PSF measured from field stars for the point source. Zhang et al.\cite{zhang2026} apply it in a particularly demanding regime, combining an extreme source colour, a faint rest-frame optical host fraction and residual stacking to reach the required signal-to-noise ratio. The tests below reproduce or modify specific elements of their analysis, so the relevant procedure is summarized here.

Their final sample of 217 objects was drawn from a parent catalogue of 434 by a blue colour cut, a magnitude limit, the visual exclusion of mergers and disturbed morphologies, and a requirement that the point-source-plus-Sérsic fit return a reduced $\chi^2$ below 2. Images were analysed in $61\times61$-pixel cutouts in F277W and F444W and $31\times31$-pixel cutouts in F115W and F150W. A global background was removed with Background2D and an additional local flat correction was applied. The fits were performed with galight version 0.2.1, which uses lenstronomy for image modelling.

Empirical PSFs were constructed with PSFEx\cite{bertin2011} from multiple unsaturated field stars in each COSMOS-Web subregion. PSFEx represented their spatial variation as a low-order polynomial of detector position, providing a local stellar PSF at the position of each LRD. Zhang et al. first fitted every object with a point-source-only model, with any neighbouring source in the cutout represented by its own Sérsic component so that its light is accounted for separately. The fitted point source and neighbours were subtracted from the observed image. The resulting residual contained everything that these models had not absorbed, including host emission, noise and systematic modelling errors. No individual residual showed evident extended emission, which Zhang et al. attributed to its low signal-to-noise ratio. They therefore aligned the 217 residuals on the LRD positions and averaged them without weighting. This initial stack showed negative central flux from over-subtraction and a positive ring-like excess in the outskirts.

For their quantitative measurement, Zhang et al. fitted the same original images a second time, now adding a Sérsic component for the LRD itself alongside those of the neighbours. The Sérsic effective radius was restricted to 0.03--0.10 arcsec, its index to 1--4, both ellipticity components to $-0.3$--0.3 and its centroid to within two pixels. The fitted point source and neighbours were subtracted while the Sérsic component of the LRD was retained, so the candidate host emission remains in the residual. That component serves to keep the point-source amplitude unbiased and to supply the alignment centroid, and its individual amplitude is not their primary reported estimator, which Zhang et al. attribute to the difficulty of establishing an extended component in any single object at this depth. These residuals were shifted by integer numbers of pixels without interpolation to align their fitted Sérsic centroids, and were then averaged. A single Sérsic profile was fitted to the resulting stack.

Zhang et al. supported their interpretation with controls using stars, compact galaxies, simulated sources and alternative empirical PSFs. As a further check, they aligned and stacked the original images before decomposition and found the width of that stack to lie between the stellar PSF and the residual stack.

The resulting Sérsic component contains 39.6\%, 33.9\%, 23.2\% and 11.8\% of the averaged total flux in F115W, F150W, F277W and F444W. At the mean redshift of 6.45, F444W samples light emitted near $0.60\,\mu\mathrm{m}$ and was therefore used to characterize the putative host in the rest-frame optical. Its stacked component has a half-light radius of approximately 210 parsecs and an apparent magnitude of 27.71 AB. Combining its flux with the three shorter-wavelength measurements in a four-band spectral fit spanning the rest-frame $4000\,\text{\AA}$ break yielded a stellar mass of $10^{9.02}\,M_\odot$.

\subsection{Representative nuclear spectrum and zero-host test}\label{sec:zero-host-test}

The first test asks whether the point-source-plus-Sérsic decomposition of Zhang et al., \hyperref[sec:nucleus-host-decomposition]{summarized in the previous subsection}, assigns a Sérsic component to an image that contains only an unresolved LRD nucleus. The calculation proceeded in five steps: (1) a representative nuclear spectrum was derived from the fluxes Zhang et al. publish for each object; (2) two model P-PSFs were computed with STPSF, one for that spectrum and one for a mixture of stellar spectra; (3) the difference between those two models was transferred onto the empirical stellar PSF released by Zhang et al., giving a source-conditioned LRD P-PSF; (4) that P-PSF was scaled to a total flux, producing a synthetic image containing one unresolved source and no host galaxy; and (5) that image was fitted with the empirical stellar PSF and the bounded Sérsic family. Any Sérsic flux recovered can then come only from the difference between the LRD and stellar P-PSFs.

The nuclear spectrum was calculated from the object-level catalogue released by Zhang et al.\cite{zhang2026}. For object $j$ in filter $k$, let $t_{jk}$ and $h_{jk}$ be their published total and host fluxes. Both are fluxes integrated over the corresponding fitted model components. The flux assigned by their decomposition to the unresolved nucleus is
\begin{equation}
p_{jk} = t_{jk} - h_{jk}.
\label{eq:nuclear_flux}
\end{equation}
The logarithmic spectral calculation below requires a positive $p_{jk}$ in every filter. Four of the 217 catalogue objects fail that requirement. Object 274713 has $p_{jk}$ less than or equal to zero in F150W, objects 293713 and 307187 in F115W, and object 381683 in F277W. These four objects comprise 1.8\% of the sample and were excluded before any median or slope was calculated, under a rule that made no reference to their colours or fitted host fractions. For each filter, the representative nuclear flux is the median of the remaining 213 values
\begin{equation}
\tilde{p}_k = \operatorname*{median}_{j} p_{jk}.
\label{eq:median_flux}
\end{equation}
The four values $\tilde{p}_k$ in F115W, F150W, F277W and F444W define the representative LRD nuclear spectrum. Their common normalization has no effect because every P-PSF is normalized to unit integrated flux.

An effective power-law slope $\beta_k$ was inferred from the four median nuclear fluxes and used to approximate the in-band LRD spectrum supplied to STPSF in each filter, where $\lambda_k$ is the pivot wavelength of filter $k$. For the two inner filters, $k=2,3$, it is
\begin{equation}
\beta_k = \frac{\ln \tilde{p}_{k+1} - \ln \tilde{p}_{k-1}}{\ln \lambda_{k+1} - \ln \lambda_{k-1}}.
\label{eq:slope_inner}
\end{equation}
The two endpoints require one-sided differences
\begin{equation}
\beta_1 = \frac{\ln \tilde{p}_2 - \ln \tilde{p}_1}{\ln \lambda_2 - \ln \lambda_1}, \qquad
\beta_4 = \frac{\ln \tilde{p}_4 - \ln \tilde{p}_3}{\ln \lambda_4 - \ln \lambda_3}.
\label{eq:slope_endpoints}
\end{equation}
This gives $\beta$ values of $-0.364$, $+0.879$, $+2.562$ and $+4.125$ from F115W to F444W.

The primary F444W slope uses the F277W and F444W points because F444W is the final point of the spectrum. Dependence on that endpoint choice was tested by fitting quadratic and cubic curves to all four median points in the same log-flux versus log-wavelength plane and evaluating their gradients at F444W. The three F444W slope estimates were then carried separately through the P-PSF generation and Sérsic fit described below. The quadratic and cubic slopes are $+5.130$ and $+5.471$ and produce Sérsic fractions of 21.96\% and 23.05\%. The adjacent-pair slope of $+4.125$ produces 18.73\% and is therefore the most conservative of the three F444W estimates. These slopes are photometric interpolation models, so the \hyperref[sec:robustness]{alternative continua and emission-line spectra of the robustness test} quantify the dependence on this assumption.

For each filter $k$, STPSF version 2.2.0\cite{perrin2014,rieke2023} calculated one model P-PSF for the representative LRD spectrum and six model P-PSFs for stellar spectra. Using $P_k[f]$ to denote the normalized P-PSF returned for source spectrum $f$ gives
\begin{equation}
P^{\mathrm{sim}}_{\mathrm{LRD},k} = P_k\!\left[\lambda^{\beta_k}\right], \qquad
P^{\mathrm{sim}}_{\mathrm{star},k} = \frac{1}{6}\sum_T P_k\!\left[B_\nu(T)\right],
\label{eq:model_ppsfs}
\end{equation}
where $B_\nu(T)$ is the Planck spectrum per unit frequency at temperature $T$, and $T$ takes the values 3,000, 4,000, 5,000, 6,000, 8,000 and 10,000~K. The second expression averages the six P-PSF images with equal weights and provides the stellar reference used in the transfer. It approximates the temperature range represented in the empirical template, while \hyperref[sec:robustness]{alternative stellar spectra are tested in the robustness test}. Both calculations use the same telescope pupil, filter, detector and sampling. Their difference is consequently produced by the different spectral weights within the filter.

STPSF evaluates the wavelength integral across each filter with 25 wavelength samples. Each sample carries the weight set by the filter throughput and the chosen source spectrum. The optical calculation used the segmented JWST pupil, fourfold internal oversampling, a 201 by 201 pixel output and the detector-distorted image sampled at 0.03 arcsec per pixel. The 201-pixel output covers 6.03 arcsec and retains the diffraction wings before the fitting region is cut out. NRCA1 was used for F115W and F150W and NRCA5 for F277W and F444W. Every calculated P-PSF was normalized to unit integrated flux.

The adequacy of the 25-sample integration was verified by repeating the F444W calculation with 50 and 100 samples. For the representative nuclear spectrum the allocation is 18.7274\%, 18.7305\% and 18.7314\% with 25, 50 and 100 samples. For the strongest narrow-line case tested, an H$\alpha$ line carrying 40\% of the in-band flux, the corresponding values are 40.0311\%, 40.0003\% and 40.0014\%. The largest departure from the 100-sample result is 0.030 percentage points, so 25 samples are converged for smooth continua and for strong lines alike.

The nuclear template must carry the LRD spectrum and the measured detector structure at the same time, and each of those comes from a different source. The two STPSF images supply the source-dependent optical change. The empirical stellar P-PSF $v_k$ released by Zhang et al. supplies the detector response, charge diffusion, resampling and image-processing structure measured in their data and was normalized to unit integrated flux before the transfer. A hat denotes a Fourier transform and an asterisk denotes complex conjugation. The regularized relative change $\Delta_k$ and the source-conditioned empirical LRD P-PSF $u_k$ were calculated as
\begin{equation}
\hat{\Delta}_k = \frac{\left(\hat{P}^{\mathrm{sim}}_{\mathrm{LRD},k} - \hat{P}^{\mathrm{sim}}_{\mathrm{star},k}\right)\left(\hat{P}^{\mathrm{sim}}_{\mathrm{star},k}\right)^{\!\ast}}{\left|\hat{P}^{\mathrm{sim}}_{\mathrm{star},k}\right|^{2} + \varepsilon},
\label{eq:relative_change}
\end{equation}
\begin{equation}
\hat{u}_k = \hat{v}_k\left(1 + \hat{\Delta}_k\right).
\label{eq:conditioned_ppsf}
\end{equation}
Multiplication by the Fourier transform of the empirical stellar template transfers the relative optical change to the measured PSF, and the resulting source-conditioned P-PSF was normalized to unit integrated flux. The regularization parameter $\varepsilon$ was set to $3 \times 10^{-4}$ of the peak power in the simulated stellar P-PSF, and varying it from $10^{-4}$ to $10^{-3}$ changed the F444W Sérsic allocation by less than one percentage point. Identical simulated spectra make the relative change zero and return $u_k = v_k$, an identity recovered here to a maximum absolute pixel error of $7 \times 10^{-18}$.

Whatever Sérsic flux the fit later assigns has to be attributable to the template mismatch alone, so the image was built to contain nothing else. The synthetic zero-host image in filter $k$ was then defined by
\begin{equation}
y_k = F_k u_k.
\label{eq:zero_host_image}
\end{equation}
The scalar $F_k$ is the total flux represented by the published stack. It was recovered by dividing the summed flux of the published stacked Sérsic model by its published Sérsic-to-total fraction, although in this noise-free linear test $F_k$ only sets the image brightness and leaves the recovered fraction unchanged. Every pixel in $y_k$ belongs to the single unresolved P-PSF $u_k$, so the input Sérsic flux is exactly zero.

The image was fitted with the same point-source-plus-Sérsic model family and published parameter bounds, using the numerical implementation described below. Each zero-host image was fitted with the point-source-plus-Sérsic model
\begin{equation}
\begin{aligned}
m_k(\boldsymbol{x}) ={}&
a_k v_k(\boldsymbol{x})
+ c_{x,k}\,\partial_x v_k(\boldsymbol{x})
+ c_{y,k}\,\partial_y v_k(\boldsymbol{x}) \\
&+ b_k
\left[
S(R_{\mathrm{e}},n,e_1,e_2;\boldsymbol{x}_{S,k})
\circledast v_k
\right](\boldsymbol{x})
+ B_k(\boldsymbol{x}),
\qquad b_k \ge 0.
\end{aligned}
\label{eq:fit_model}
\end{equation}
The first term is the empirical stellar P-PSF fitted as the nucleus. The second is a Sérsic profile convolved with that P-PSF, and the symbol $\circledast$ denotes convolution. The coefficients $c_{x,k}$ and $c_{y,k}$ represent first-order shifts of the point-source centre, while $\boldsymbol{x}_{S,k}$ is the fitted Sérsic centroid. The background $B_k$ contains a constant and one gradient along each image axis. Both $v_k$ and every convolved Sérsic template were normalized to unit integrated flux, so $a_k$ and $b_k$ are their fitted fluxes. The Sérsic bank discretizes the bounds published by Zhang et al., an effective radius of 0.03 to 0.10 arcsec, an index of 1 to 4 and ellipticity components between $-0.3$ and $+0.3$, into 29 radii, 13 indices and five values of each ellipticity component, giving 9,425 intrinsic shapes, with the centre free within their two-pixel interval. Every shape was first evaluated on a one-pixel centre grid across that interval, the complete bank was then evaluated on a 0.25-pixel grid around the best coarse cell, and the 100 leading combinations of shape and centre were refined continuously. For every trial, the point amplitude, first-order shifts of the point centre, non-negative Sérsic amplitude and background plane were solved together. The model and its bounds are those published by Zhang et al., while the discretized bank and the first-order centre shift are a numerical route to solving it and not the optimizer those authors used. The consequence of that difference is measured directly in the \hyperref[sec:galight-control]{independent galight fit}.

Zhang et al. define the released normalized residual as $r_i=(d_i-m_i^{\mathrm{pub}})/\sigma_i$, where $d_i$ and $m_i^{\mathrm{pub}}$ are the published stacked data and model. The uncertainty of pixel $i$ was therefore recovered as
\begin{equation}
\sigma_i = \frac{\bigl|d_i - m_i^{\mathrm{pub}}\bigr|}{\bigl|r_i\bigr|}.
\label{eq:pixel_uncertainty}
\end{equation}
Ratios were evaluated only where the absolute normalized residual exceeded 0.03 and the absolute data--model difference exceeded $10^{-10}$. Only finite positive ratios were retained. Unstable or non-positive values were replaced by the median valid uncertainty in that band, and each map was clipped at its 0.5th and 99.5th percentiles. All 709 pixels entering each fit consequently had finite positive uncertainties. Varying the residual threshold from 0.01 to 0.10 and the clipping from none to the 1st and 99th percentiles moved the F444W allocation only from 18.47\% to 19.04\%.

The fit selected the Sérsic profile with the smallest noise-weighted squared residual
\begin{equation}
\chi^2 = \sum_i \left(\frac{y_{k,i} - m_{k,i}}{\sigma_i}\right)^{2}.
\label{eq:chi_square}
\end{equation}
The zero-host allocation for that minimum-$\chi^2$ fit is
\begin{equation}
A_k = \frac{b_k}{F_k}.
\label{eq:allocation}
\end{equation}
The resulting allocations are given in Table~\ref{tab1}. The amount of P-PSF light physically redistributed by the spectral change was measured independently as
\begin{equation}
R_k = \frac{1}{2}\sum_i \bigl|u_{k,i} - v_{k,i}\bigr|.
\label{eq:redistributed_light}
\end{equation}
The factor one half avoids counting each transfer twice. The redistributed-light fractions in each filter are given in Table~\ref{tab1}. The larger Sérsic allocations arise because the fit simultaneously lowers the point-source amplitude and raises the Sérsic amplitude. A small change in the normalized P-PSF can therefore move a much larger fitted flux between the two model components.

In F115W, the LRD P-PSF places slightly more light in the centre than the stellar template (Fig.~\ref{fig1}b) and has a negative projection onto every allowed Sérsic profile. The constraint $b_k \ge 0$ consequently fixes the allocation at zero.

Three implementation checks were applied to the primary result. The common fit used a 31 $\times$ 31 pixel crop and the pixels within 0.45 arcsec of its centre. Repeating F444W on its 61 $\times$ 61 pixel crop changed the allocation by 0.06 percentage points. Fitting the complete synthetic point and fitting only its difference from $v_k$ returned the same Sérsic flux to one part in $10^{15}$. A direct fit to the STPSF difference before empirical transfer preserved the wavelength ordering and the exact zero in F115W.

The four allocations, the redistributed-light fractions, the Sérsic shape and centre selected in each band and the F444W stability series are released in the Table\_1, Sersic\_centre\_search and F444W\_stability worksheets of the Source Data.

\subsection{Robustness to the source spectrum, detector position and stellar reference}\label{sec:robustness}

The preceding section reports one number obtained with one nuclear spectrum, one stellar reference and one detector position. This section asks how far that number moves when each of those three choices is replaced. Three tests were run. Within each test, the named physical input is varied while the remaining settings of that test and the Sérsic-bank fit stay fixed. The first replaces the representative nuclear spectrum by six alternatives. The second moves the calculation to ten positions across two detectors. The third replaces the blackbody stellar reference by model atmospheres. All three are reported for F444W, the band that carries the published size, luminosity and stellar mass. Every allocation obtained is plotted in Extended Data Fig.~\ref{edfig1}, and the values are released in the ED1\_spectra, ED1\_stellar\_reference, ED1\_positions and Atmosphere\_models worksheets of the Source Data.

The first test isolates the dependence on the assumed nuclear spectrum, which is the input carrying the largest uncertainty because it is inferred from the published photometry. Six alternative LRD spectra were compared against a single 4,000~K stellar reference, so that the source spectrum is the only quantity that changes. The influence of the stellar reference itself is the subject of the third test. Three of the six were smooth continua with $\beta$ of 3.0, at the blue limit of the LRD selection, 3.896, the value implied by the point-source photometry published by Zhang et al., and 5.0, a red extreme. One contained the [O\,\textsc{iii}] line at rest wavelength 500.7 nm and redshift 7.8 carrying 20\% of the F444W light. Two contained H$\alpha$ at rest wavelength 656.3 nm and redshift 6.45 carrying 20\% and 40\% of the in-band light. The six sample a broad range of continuum slopes and line-dominated spectra, placing [O\,\textsc{iii}] near the centre of F444W and H$\alpha$ near its red edge.

Their minimum-$\chi^2$ Sérsic allocations, meaning the fraction of the total light that the bounded Sérsic family absorbs from a source containing no galaxy, were 15.67\%, 18.78\%, 22.56\%, 16.15\%, 28.45\% and 39.81\%. The smallest of the six is 1.33 times the 11.8\% component published by Zhang et al. and the largest is 3.37 times it. The spread of a factor of 2.5 is dominated by the two H$\alpha$ cases, which shows that the allocation grows as the spectrum departs further from a stellar one. A stacked measurement averages many objects, so a line-dominated nucleus enters the stack in proportion to how common such objects are. The line cases sample the effect for individual objects of that kind.

The second test isolates the dependence on where the object falls on the focal plane, since the diffraction pattern varies across the field and Zhang et al. model that variation within their spatially varying template. The calculation was repeated at the centre and at four off-axis positions on each of the NRCA5 and NRCB5 detectors. STPSF regenerated both the stellar and the LRD P-PSF at every location, the local spectral change was transferred to the same empirical template and the same Sérsic bank was fitted. The ten allocations lie between 16.27\% and 20.08\%, with an equal-weight mean of 18.59\%. Within this optical-position test, detector position moves the allocation across a range of 3.8 percentage points, one fifth of its value, while reaching the published component from that mean would require a downward shift of 6.8 points.

The third test isolates the dependence on the stellar reference itself. A blackbody is an idealization, and a real stellar atmosphere could narrow the gap to the LRD. Four solar-composition Castelli and Kurucz ATLAS9 atmospheres\cite{castelli2004} were therefore used in F444W, at effective temperatures of 4,000, 5,000, 6,000 and 8,000~K and at a main-sequence surface gravity of $\log g = 4.5$. The tabulated flux per unit frequency of each model was passed through STPSF and the four resulting P-PSF images were averaged, mirroring the averaging applied to the blackbody mixture. A diagnostic straight-line fit of $\log f_\nu$ against $\log \lambda$, made only where the F444W transmission exceeded 1\% of its maximum, gives slopes of $-2.388$, $-2.206$, $-1.949$ and $-1.866$ with a mean of $-2.102$, against $-1.642$ for the blackbody mixture and $+4.125$ for the representative LRD. Real atmospheres are therefore bluer than blackbodies of the same temperature near 4.4~$\mu$m, which widens the separation from the LRD by 0.46 in $\beta$ and raises the allocation from 18.73\% to 20.31\%. This replacement changes the temperature coverage from 3,000--10,000~K to 4,000--8,000~K along with the spectral physics, so the 1.58 points measure the two together. The blackbody reference is retained for the primary result because it is the choice that works against the conclusion drawn here.

The three tests together measure how far each uncertainty can move the F444W allocation. The assumed nuclear spectrum moves it between 15.67\% and 39.81\%, detector position moves it across 3.8 percentage points around a mean of 18.59\%, and the stellar reference moves it upwards by 1.58 percentage points. The lowest value reached anywhere in the three tests is 15.67\%, which is 1.33 times the component published by Zhang et al. None of these uncertainties considered individually brings the F444W allocation down to 11.8\%, and the two replacements with the strongest physical justification, a real stellar atmosphere in place of a blackbody and an H$\alpha$-bearing nucleus in place of a smooth continuum, both move it upwards.

One further freedom belongs to the fit itself. Zhang et al. allow the Sérsic centroid to move within two pixels, so the allocation could in principle depend on where inside that interval the fit settles. The F444W construction was therefore repeated on a grid of 81 centroid positions covering the whole interval in half-pixel steps. Every position within $\Delta\chi^2 = 9$ of the best fit returns an allocation of at least 14.77\%, which is 1.25 times the published component. No near-minimum solution on this grid brings the F444W allocation down to 11.8\%.

The complete grid is released in the Sersic\_centre\_degeneracy worksheet of the Source Data.

\subsection{Paired zero-host injections on real COSMOS-Web backgrounds}\label{sec:paired-injections}

The noise-free calculation establishes the susceptibility of the decomposition. The next test asks how much paired chromatic allocation is recovered on selected real COSMOS-Web backgrounds at realistic source brightness and photon noise. The test that imposes them is a paired injection. Two artificial point sources that differ only in their spectrum are placed one at a time on the same patch of real sky, both are fitted with the stellar template released by Zhang et al., and the Sérsic allocation given to one is subtracted from the Sérsic allocation given to the other. One hundred such pairs were built and measured in five steps: (1) real sky and real brightnesses are taken from the COSMOS-Web release; (2) an empty patch is located inside each cutout; (3) the two members of a pair are constructed and injected; (4) both are fitted with that stellar template and the bounded Sérsic family; and (5) the paired difference is formed and its uncertainty is estimated. A difference consistent with zero would show that a change of spectrum alone produces no measurable extended component under these conditions. A difference reliably above zero would show that it does, and its size is what the spectrum alone contributes on this sky.

The COSMOS-Web parent catalogue of 434 LRDs was cut to the sky area of the B9 F444W mosaic\cite{casey2023} and then to objects with usable photometry, meaning a finite F115W and F150W magnitude above zero, a colour $m_{115}$ minus $m_{150}$ below 0.8 mag, which is the blue cut applied by Zhang et al., and a finite F444W magnitude between 0 and 30. Twenty-five objects in that tile satisfy all three conditions, and each supplied a 141 by 141 pixel cutout centred on its catalogue position. The sample stops one step short of the selection made by Zhang et al., who additionally required a reduced $\chi^2$ below 2 from their own galight fits. That quantity belongs to their fitting run and does not form part of the released catalogue, so the final 217-object sample is not reproducible from public data alone. The consequence is contained, because these positions are used only to supply realistic brightnesses and realistic sky, and every conclusion below concerns the difference between two artificial sources placed on that sky.

An injection landing on a real object would measure blending, so a fixed set of candidate patch centres was examined systematically in every cutout and each candidate was scored for cleanliness. Every cutout yielded acceptable candidates, so the set was never extended. The twelve centres, identical for every cutout, sit at pixel offsets $(\pm 35, 0)$, $(0, \pm 35)$, $(\pm 32, \pm 32)$, $(\pm 40, \pm 20)$ from the catalogue position. Each candidate was cropped to 61 by 61 pixels. Its background level was defined as the median inside an annulus between 8 and 15 pixels from the candidate centre, and its noise $\sigma$ as 1.4826 times the median absolute deviation in that same annulus, the factor 1.4826 being the constant that turns a median absolute deviation into the standard deviation of a Gaussian distribution. A pixel counts as contaminated when it departs from the background level by more than 5$\sigma$, and the contaminated set is then grown by one pixel so that the wings of a neighbour also count. The score adds three terms, the offset of the median inside 4 pixels from the background level in units of $\sigma$, one hundred times the contaminated fraction inside 4 pixels, and ten times the contaminated fraction in the annulus, so that a lower score marks a cleaner patch. The four lowest-scoring patch centres were kept for each of the 25 catalogue positions, giving 100 patches and therefore 100 pairs. Because those patches were selected for cleanliness, this experiment measures the chromatic term on clean real sky, and the full neighbour environment surrounding an LRD is outside what it tests.

Each pair receives two artificial point sources injected one at a time into its patch. One carries the empirical stellar P-PSF released by Zhang et al. The other carries the source-conditioned LRD P-PSF of equation~\ref{eq:conditioned_ppsf}, which is that same empirical template transferred to the representative LRD nuclear spectrum of in-band slope $\beta = +4.125$ in F444W. Both are pure unresolved points and neither contains any host galaxy. Everything except the spectrum is held identical. Both are shifted to the same subpixel phase, drawn once per pair from a uniform distribution on the interval from $-0.5$ to $+0.5$ pixels in each axis. Both are scaled to the same catalogue F444W magnitude of that pair's parent position, converted to the surface-brightness units of the mosaic by
\begin{equation}
F = \frac{3631\,\mathrm{Jy} \times 10^{-0.4\,m_{444}}}{10^{6}\,\Omega_{\mathrm{pix}}},
\label{eq:flux_conversion}
\end{equation}
where $\Omega_{\mathrm{pix}}$ is the solid angle of one pixel, $2.1154 \times 10^{-14}$ sr for this mosaic. Photon noise is then added to each of the two injections separately, accounting for the different light concentration in each case. Writing $S$ for the surface brightness of an injected point in one pixel and $S_{+}=\max(S,0)$, the expected number of detected counts in that pixel is $S_{+}k$, where $k$ converts one unit of mosaic surface brightness into counts,
\begin{equation}
k = 5.0008 \times 10^{4}\ \text{counts per}\ \mathrm{MJy\,sr^{-1}},
\label{eq:count_conversion}
\end{equation}
for this mosaic. That value combines its 18,553.104 s exposure time with the flux calibration recorded with the data, which converts a count rate of one count per second into 8.7201 microjansky per square arcsecond. The Poisson dispersion of the counts is their square root, so the dispersion in the units of the mosaic is
\begin{equation}
\sigma = \sqrt{S_{+}/k}.
\label{eq:photon_noise}
\end{equation}
A single field of standard normal deviates is drawn once per pair and multiplied by the corresponding per-pixel dispersion for each injection. This common-random-number design reduces Monte Carlo variance while retaining the different photon-noise amplitudes implied by the two P-PSFs. The phases and the noise fields come from the NumPy PCG64 generator seeded once with 240826 at the start of the run, and they are drawn in pair order.

Both injections are then fitted with a point component carrying the empirical stellar P-PSF released by Zhang et al. and a Sérsic component drawn from the complete bank of 9,425 profiles inside the bounds those authors published, which is the model of equation~\ref{eq:fit_model} with the Sérsic centre held at the injected position. The fitting domain is the uncontaminated area of the patch inside 0.45 arcsec of the injected centre, and the per-pixel weight is the robust dispersion measured in the 8 to 15 pixel annulus of that patch. Once the minimum-$\chi^2$ member of the bank has been selected, the centre of the point component is refined nonlinearly within one pixel, so that a residual subpixel centring error cannot present itself as extended light.

Background structure, the imperfect match between a template centred on the pixel grid and a source at an arbitrary subpixel position, and a Sérsic amplitude that cannot fall below zero all contribute, so even the stellar injection, whose spectrum matches the template exactly, is assigned some Sérsic flux. Quoting the LRD value alone would therefore overstate the chromatic term, and the quantity formed for each pair is the difference between the two Sérsic allocations measured on the same patch,
\begin{equation}
D =
\frac{b_{\mathrm{LRD}}}{a_{\mathrm{LRD}}+b_{\mathrm{LRD}}}
-
\frac{b_{\mathrm{star}}}{a_{\mathrm{star}}+b_{\mathrm{star}}},
\label{eq:paired_difference}
\end{equation}
where $a$ and $b$ are the fitted point-source and Sérsic fluxes. Because the spectrum is the only experimental factor changed within a pair, $D$ estimates its effect conditional on the shared background, phase, brightness and random field. Pairing reduces their influence without cancelling it algebraically.

Across the 100 pairs the median Sérsic allocation is 8.8\% for the stellar injection and 24.9\% for the LRD injection, and the median of the 100 differences is 13.7 percentage points. The 8.8\% value is the implementation-specific null allocation in this individual-patch experiment and is not interpreted as physical host light. The paired difference of 13.7 percentage points is the primary estimand and is comparable in size to the complete 11.8\% component that Zhang et al. attribute to a host galaxy in this band.

The uncertainty on that median was estimated by rebuilding the sample 10,000 times. Each rebuild draws 25 catalogue positions at random from the original 25, allowing the same position to be drawn more than once, and takes all four of its patches whenever a position is drawn, because four pairs that share a position also share its brightness and its surrounding sky and are not independent. Recomputing the median on every rebuild gives a 95\% interval from 9.0 to 16.6 percentage points. None of the 10,000 rebuilt samples places the median at or below zero, and the difference is positive in 92 of the 100 individual pairs. The effect is therefore present in almost every pair and not carried by a few outliers. 

One further split separates a deterministic effect from a noise artefact. The eight positions brighter than 25th magnitude in F444W contribute 32 pairs, and their median difference is 17.3 percentage points with a 95\% interval from 16.8 to 17.8. Brighter sources carry proportionally less photon noise, so a deterministic mismatch should emerge more clearly among them while a difference manufactured by noise should fade. The measured value rises, which is consistent with a deterministic mismatch and argues against a purely noise-generated difference. 

The paired experiment therefore shows that a chromatic excess comparable to the published 11.8\% component survives on selected clean COSMOS-Web backgrounds. The 13.7-point paired estimate should not be interpreted as a direct reduction from the noise-free 18.73\%, because the two calculations use different estimands and fitting conditions.

The 25 catalogue positions, the 100 pair records and the resulting estimands are released in the Paired\_positions, Paired\_injections and Paired\_estimands worksheets of the Source Data.

\subsection{Floor of the stacked estimator on spatially distributed blank sky}\label{sec:stacked-floor}

Averaging 217 residuals is what makes a faint extended component measurable at this depth, and it is also the obvious objection to every per-object measurement reported so far, since a systematic seen in individual fits could be expected to wash out of a stack. An average separates two kinds of contribution. What is uncorrelated between objects falls as the square root of the sample size, while what every object shares does not fall at all. This test asks which of the two a chromatic mismatch is, and therefore whether stacking removes the per-object chromatic excess or concentrates it. It proceeds in five steps: (1) 217 patches of blank sky distributed across the mosaic are extracted; (2) three populations are built on them, differing only in what is injected; (3) every image is fitted and its point component and background subtracted; (4) the 217 residuals of each population are stacked and fitted with a single Sérsic; and (5) the three stacked allocations are compared.

Patches drawn from one neighbourhood share their background, so positions distributed across the mosaic were required. Four hundred and twenty positions were drawn at random across the B9 F444W footprint, of which 245 fell on exposed sky. Each was cropped to 61 by 61 pixels and scored by the rule used for the paired injections, and the 217 cleanest were retained, one for each object in the published sample. Each patch received one magnitude drawn at random from the 434 valid F444W magnitudes of the parent catalogue, so that the brightness distribution is that of the population.

Three populations were propagated through one identical pipeline, two of them to isolate the spectrum and the third to establish what the sky alone contributes. The null contains no injected source. The stellar population contains an unresolved point carrying the empirical stellar P-PSF, whose spectrum matches the fitting template exactly. The chromatic population contains an unresolved point carrying the source-conditioned LRD P-PSF of equation~\eqref{eq:conditioned_ppsf}. None of the three contains a host galaxy, and all three share the same patches, magnitudes, subpixel phases and random field. Every image was fitted with the empirical stellar P-PSF and the Sérsic bank. The fitted point and background were subtracted, each residual was registered to a common subpixel phase, and the 217 residuals were averaged without weighting before fitting a Sérsic profile over the published two-pixel centroid interval. This subpixel registration differs from the integer-pixel translations used by Zhang et al. and tests the fully coherent limit.

The blank-sky stack returns 0.720\% of the reference flux, with $\Delta\chi^2=18.8$, showing that the selected sky contributes little compared with the published 11.8\% component. The stellar stack returns 9.48\%, whereas an otherwise matched control with every stellar source placed exactly on the pixel grid returns 0.00\%. Its absolute allocation is therefore dominated by the subpixel representation of the point source in this first-order fitting implementation.

The chromatic stack returns 38.49\%, compared with 9.48\% for the stellar stack, giving a difference of 29.0 percentage points. Their respective improvements are $\Delta\chi^2=3{,}263$ and 460. Because spectrum is the only injected property that differs between the two populations, their contrast measures a spectrum-dependent residual in this implementation. The 29.0-point stacked contrast and the 13.7-point median paired contrast are different estimands, with different normalization, centring and aggregation, and should not be compared numerically. The relevant result is that the chromatic residual remains coherent when the sky is averaged down. A significant stacked component is therefore not by itself evidence of astrophysical extension.

Every injection carries the same representative nuclear spectrum, so the 29.0-point result describes a deliberately coherent stress test. A real distribution of slopes, lines and redshifts can weaken or strengthen the stacked residual. The \hyperref[sec:catalogue-estimate]{catalogue-level calculation}, which assigns each object its own measured slope, provides the population-level estimate. The three stacks and the patch positions are released in the Stacked\_floor and Stacked\_floor\_patches worksheets of the Source Data.

\subsection{Independent fit with galight}\label{sec:galight-control}

The preceding allocations were obtained with a discretized Sérsic bank and a first-order expansion of the point-source position, and not with the optimizer used by Zhang et al. To test whether the result depends on that implementation, the brightest B9 validation source was fitted independently in the software those authors used. The test proceeds in four steps: (1) the brightest validation object is selected from the B9 sample; (2) three hostless configurations are built on one real background patch; (3) all three are fitted by galight under the bounds published by Zhang et al.; and (4) the recovered Sérsic fractions are compared.

The software is galight version 0.2.1 and lenstronomy version 1.11.10\cite{ding2021,birrer2021}. The object is catalogue entry 845096, whose F444W magnitude of 22.760 corresponds to an injected flux of 135.128 in the surface-brightness units of the mosaic, placed on its cleanest B9 background patch with the same background and uncertainty map in all three configurations.

Three hostless configurations were fitted. The matched configuration injects a stellar point and fits it with the stellar template. The mismatched configuration changes only the injected spectrum, replacing the stellar point with an LRD point while retaining the stellar fitting template. The corrected configuration retains that LRD point and replaces the fitting template with its source-conditioned P-PSF. Comparison of the first two therefore measures the chromatic allocation, and comparison of the last two measures its correction.

Each galight model contains one point source and one elliptical Sérsic component. The Sérsic effective radius is restricted to 0.03--0.10 arcsec, its index to 1--4, both ellipticity components to $-0.3$ and $+0.3$ and its centre to $\pm 0.06$ arcsec, matching the two-pixel interval used by Zhang et al. The point-source centre is free within two detector pixels and is optimized nonlinearly, so the first-order treatment of the point centre used in the Sérsic-bank fits is absent here. Particle-swarm optimization uses 50 particles and 50 iterations in two consecutive passes, with a supersampling factor of two.

The recovered Sérsic fractions are 0.036\% in the matched configuration, 13.97\% in the mismatched configuration and 0.0048\% in the corrected configuration. The chromatic excess is therefore 13.93 percentage points and is removed when the fitting P-PSF is conditioned on the source spectrum. Between the matched and mismatched configurations the fitted point loses 18.01 flux units while the Sérsic component gains 17.87, which shows the exchange of flux between the two fitted components directly.

The corresponding reduced chi-squares are 1.103, 1.315 and 1.102. The mismatched fit is degraded but remains below the reduced chi-square threshold of 2 used by Zhang et al. for sample selection. Their quality criterion would therefore retain this hostless source despite assigning 13.97\% of its light to a Sérsic component. This agrees with the finding of Zhuang and Shen that fit quality need not reveal PSF mismatch in compact-host decompositions\cite{zhuang2024}.

This single bright case shows that the chromatic allocation is reproduced by the original nonlinear fitting software, while the 100 paired trials determine its population-level uncertainty. Each fraction is normalized by the total fitted point-source-plus-Sérsic flux, as in equation~\eqref{eq:paired_difference}, and is therefore comparable with the 13.7-percentage-point paired contrast and not with the allocations defined by equation~\eqref{eq:allocation}. The three fits are released in the Galight\_control worksheet of the Source Data.

\subsection{Source-conditioned correction and host preservation}\label{sec:correction}

The remedy tested here is to build the nuclear template from the spectrum of the source being fitted. In this controlled experiment the input spectrum is known, so the true unresolved pattern lies inside the reduced model and the condition of equation~\eqref{eq:false_allocation} is satisfied exactly. A template that also absorbed genuine extended light would leave the measurement no better than the one it replaces, so removing the spurious component is only half of what has to be shown. This test therefore injects a galaxy of known brightness and known shape and asks what each nuclear template returns for it. It proceeds in four steps: (1) a synthetic image is built containing an unresolved nucleus carrying the source-conditioned LRD P-PSF together with an injected exponential galaxy of known flux and known effective radius, placed on a real background patch; (2) a matched control image is built in which that nucleus is replaced by one carrying the empirical stellar P-PSF, with the same galaxy, the same background, the same subpixel position and the same noise; (3) the first image is fitted twice, once with the empirical stellar P-PSF as the nuclear template, which is the conventional procedure, and once with the source-conditioned P-PSF, which is the correction; and (4) the galaxy flux recovered in each of those two fits is compared with the galaxy flux recovered from the matched control, so that all differences arise from how the nuclear P-PSF is modelled.

The injected galaxy is a circular exponential profile of Sérsic index 1, added around the chromatic nucleus at three brightnesses, 5\%, 12\% and 25\% of the total flux, and three effective radii, 0.030, 0.0375 and 0.050 arcsec, which gives nine combinations of brightness and size. Each combination is measured on 32 trials, drawn from eight objects brighter than F444W magnitude 25 with four background patches each. The injected galaxy is convolved with the same empirical stellar P-PSF used for the fitted Sérsic component, making its spatial model exact by construction and isolating errors caused by the nuclear P-PSF. The amplitudes and the centres of the nucleus and of the galaxy are fitted independently, together with a background plane. The matched control holds the identical galaxy, the same background patch, the same subpixel position and the same noise draw, and differs only in the P-PSF carried by its nucleus, so the comparison isolates the modelling of the nuclear P-PSF only. The shape of the injected galaxy is supplied to both fits, which makes this a measurement of how much of its flux survives at a known shape.

Fitting the chromatic nucleus with the conventional stellar template returns more galaxy flux than the matched control in all nine combinations, by between 7.70 and 17.09 percentage points of the total (Extended Data Table~\ref{tabED1}). This excess is nuclear light assigned to the Sérsic component. Its size is comparable to the entire 11.8\% component published in this band, showing that the mismatch can both create a false host component and inflate the recovered flux of a genuine host. The absolute excess is largest for the most compact profiles, while its importance relative to the host is greatest for the faintest injected hosts. A galaxy carrying 5\% of the total light at an effective radius of 0.030 arcsec is recovered at a median of 18.3\%, against 3.8\% for its matched control, while one carrying 25\% at 0.050 arcsec is recovered at 30.7\% against 22.9\%. Fitting the same image with the source-conditioned template instead moves the recovered fraction away from its control by at most 0.78 percentage points in absolute median. The matched control itself lies below the injected value in every combination, because subtracting the background removes some genuine galaxy light as well, so the correction is judged against that control and not against the injected truth. Measured against the injected value, the corrected fit differs by up to 7.77 percentage points in median, since the same loss applies to it.

This experiment tests flux preservation at a supplied host shape. When the Sérsic shape is instead fitted blindly, the recovered fraction can differ from the injected truth by up to 11.3 percentage points in median because the host model also exchanges flux with the real background. Absolute sizes, indices and fluxes therefore require an image-level refit and are not claimed here.

The nine injected truths, their matched controls and the resulting estimands are released in the Host\_preservation and Host\_preservation\_summary worksheets of the Source Data.

\subsection{Two-mode approximation for an unknown nuclear spectrum}\label{sec:two-mode}

Conditioning the nuclear P-PSF on the spectrum of the source requires an estimate of that spectrum, which for a real LRD is constrained by only four broad-band fluxes. This section tests whether the chromatic term can instead be represented without supplying the spectrum of each fitted source. Two fixed images derived from a library of nuclear P-PSFs are added to the model, allowing the nuclear template to adapt within the spectral family that library represents. The test proceeds in four steps: (1) two fixed images, called the diffraction modes, are derived from a library of seven nuclear P-PSFs, one stellar reference and six LRD spectra, with two of the six held out of the derivation; (2) those two modes are added as free components to the point-source-plus-Sérsic model, and the rate at which that model reports a galaxy in images containing none is measured with and without them; (3) the fraction of a genuine injected galaxy that survives the same fit is measured; and (4) the modes are applied to the \hyperref[sec:paired-injections]{100 paired injections on real sky}.

A change of nuclear spectrum changes the diffraction pattern in a specific way, and subtracting the stellar P-PSF from the P-PSF of a given LRD spectrum leaves an image showing exactly what that spectrum did, negative where light was taken out of the pattern and positive where it was put back. Those difference images, computed across a range of plausible LRD spectra, are close to scaled versions of two fixed patterns, so a weighted sum of two images should reproduce any of them. Those two images are the modes. Adding them to the model with free signed amplitudes lets the fitted point source move light between its core, its rings and its wings until it matches whatever nucleus the image actually contains, without the spectrum of that nucleus ever being supplied. All calculations in this section concern F444W, so $k$ is fixed to that filter throughout.
\begin{equation}
\begin{aligned}
m^{\mathrm{corr}}_k(\boldsymbol{x}) ={}&
a_k v_k(\boldsymbol{x})
+ c_{x,k}\,\partial_x v_k(\boldsymbol{x})
+ c_{y,k}\,\partial_y v_k(\boldsymbol{x})
+ \sum_{j=1}^{2} w_{j,k}\,\Phi_j(\boldsymbol{x}) \\
&+ b_k
\left[
S(R_{\mathrm{e}},n,e_1,e_2;\boldsymbol{x}_{S,k})
\circledast v_k
\right](\boldsymbol{x})
+ B_k(\boldsymbol{x}),
\qquad b_k \ge 0.
\end{aligned}
\label{eq:mode_model}
\end{equation}
Here $\Phi_1$ and $\Phi_2$ are the two diffraction modes and $w_{1,k}$ and $w_{2,k}$ are their fitted amplitudes, which carry no sign constraint because a change of nuclear spectrum can move light either into the core or out of it. Every other symbol keeps the meaning it has in equation~\eqref{eq:fit_model}, and the non-negativity still applies to $b_k$ alone. The two modes enter beside the point source and not beside the Sérsic component, so the freedom they add belongs to the nucleus, and the extended component is left with exactly the bounds published by Zhang et al.

The library contains seven P-PSFs in F444W, one for the stellar reference and six for LRD spectra chosen to span the cases tested here. Three are smooth continua, with $\beta$ of 3.0 at the blue limit of the LRD selection, 3.896 as implied by the published point-source photometry, and 5.0 as a red extreme. The remaining three use the $\beta = 3.896$ continuum, with [O\,\textsc{iii}] contributing 20\% of the in-band flux in one case and H$\alpha$ contributing 20\% and 40\% in the other two. Subtracting the stellar P-PSF from each of the six leaves six difference images. Two of those six, the [O\,\textsc{iii}] case and the stronger H$\alpha$ case, were set aside before anything was built, so that the basis could afterwards be tested on spectra it had never seen, and the modes were derived from the remaining four.

A singular-value decomposition of those four differences returns the patterns they have in common, ordered by how much of their variation each one accounts for. Writing $\delta_i = u_k[f_i] - v_k$ for the difference image of the $i$th training spectrum, $\Phi_1$ and $\Phi_2$ are the two leading right singular vectors of the set $\{\delta_i\}$. The first mode accounts for 99.978\% of that variation, the second for a further 0.022\%, and the third for only $1.9 \times 10^{-6}$\%. Two modes are consequently kept. The count is bounded from the other side as well, because every component added to the reduced model is one more degree of freedom competing with the Sérsic for the same light, and constraining that competition is the whole purpose of the correction.

What settles the choice is the behaviour on the two spectra the basis never saw. A single mode reconstructs 99.444\% and 99.906\% of their differential power, and two modes reconstruct 99.992\% and 100.000\%. Reconstruction is measured as $1-\lVert\delta_i-\widehat{\delta}_i\rVert_2^2/\lVert\delta_i\rVert_2^2$, where $\widehat{\delta}_i$ is the reconstruction of $\delta_i$ from the retained modes. Setting those two spectra aside is what makes this a test, because a basis derived from all six differences would describe all six by construction.

The first measurement asks how often each model reports a galaxy in an image that contains none. Controlled simulations placed a centred point source of total flux 32 on a background of per-pixel dispersion 0.008, with 500 independent noise realizations for each of the seven spectra, giving 3,500 images in which no galaxy exists. A galaxy counts as detected when the non-negative Sérsic amplitude improves $\chi^2$ by more than 2.7055, the nominal one-sided 5\% boundary for one fixed constrained amplitude. Because the fit also profiles over a bank of Sérsic shapes whose parameters are not identified under the null, the achieved false-positive rate is calibrated empirically against the matched stellar simulation and not assumed to equal 5\%. All quoted uncertainties on these rates are binomial standard errors. The conventional model, which fits every nucleus with the stellar P-PSF, reports a galaxy in 86.17\%~$\pm$~0.58\% of those images. That aggregate is diluted by the one spectrum of the seven for which the fitting template is the correct one, the stellar reference, where the rate is 3.6\%. Taken separately, each of the six LRD spectra gives a rate between 99.8\% and 100\%. Applied to a nucleus whose spectrum differs from its template, the conventional model reports a galaxy that does not exist in essentially every image, and applied to one whose spectrum matches it behaves as the threshold intends. Adding the two modes brings the rate to 4.34\%~$\pm$~0.34\%, close to the 3.6\% measured for the matched stellar reference and to the nominal 5\% boundary, with no individual spectrum outside 2.8\% to 5.8\%, and the rate is 4.36\% for the five spectra that were not withheld against 4.30\% for the two that were. The agreement between those two figures shows that the basis performs on the two withheld spectra as it does on the four from which it was derived.

The first measurement showed that the modes do not invent a galaxy where there is none. The second asks whether they destroy one that is there. The concern is concrete, because the two mode amplitudes are free in sign and nothing in the model stops them from absorbing light that belongs to a real extended component. The \hyperref[sec:correction]{injected-host test} answered that question for the source-conditioned template, which is built from a known spectrum, and it leaves it open for this approximation.

The images used here therefore contain two components. The nucleus is an unresolved point carrying the $\beta = 3.896$ continuum, and around it sits a genuine exponential galaxy of Sérsic index 1 and effective radius 0.0375 arcsec, injected at 5\%, 12\% and 25\% of the total light with 500 noise realizations of each. Brightness is the only property of the galaxy that varies here, so this measures flux preservation at one morphology, and the three effective radii of the \hyperref[sec:correction]{injected-host test} cover the dependence on size. Every image is fitted twice, once with the stellar template alone and once with the two modes added, and what is compared is the fraction of the total light each fit hands to the galaxy against the fraction that was put into it.

Fitted with the stellar template alone the recovered fractions are 18.76\%, 25.10\% and 35.74\%, so a galaxy carrying 5\% of the light is reported as carrying almost four times that. Fitted with the two modes present they are 5.29\%, 12.15\% and 24.68\%, departing from the injected values by $+0.29$, $+0.15$ and $-0.32$ percentage points. The modes return the galaxy to its true brightness across a factor of five in host fraction. Taken with the first measurement, they neither create a galaxy that does not exist nor remove one that does, and the recovery here is measured against the injected truth, not against a matched control.

The third measurement moves the modes onto real sky. Each of the \hyperref[sec:paired-injections]{100 paired injections} holds two artificial point sources placed one at a time on the same patch, one carrying the empirical stellar P-PSF and one carrying the LRD P-PSF, and the quantity formed from each pair is the difference between the fractions of fitted flux the two of them hand to the Sérsic component. That paired difference is the chromatic excess, and it is what the modes have to remove. They were added to those same fits with every other element unchanged.

Without the modes the median excess across the 100 pairs is 20.96 percentage points. With them it is $-0.035$ points, with a 95\% object-cluster interval from $-0.177$ to $+0.008$, so the excess is consistent with zero. The modes remove the chromatic excess without being given the spectrum of either injected source.

The 20.96-point baseline is not directly comparable with the \hyperref[sec:paired-injections]{13.7-point paired result}, because the two analyses treat the point centre with different estimators. The present experiment is internally controlled, with identical images, backgrounds, subpixel phases, noise draws and centre treatment before and after the modes are added, so its relevant comparison is 20.96 points without them against $-0.035$ points with them.

Two properties of that result are declared. The modes raise the absolute Sérsic allocation of both members of each pair on structured backgrounds, and that common offset cancels in the paired difference, so what this experiment measures is the removal of the chromatic term and not an absolute host flux. Separately, a subpixel displacement is itself a small change in the shape of the point pattern, so on real sky the modes could in principle absorb centring residuals alongside chromatic ones. The 3,500 point-only simulations bear on that directly, because their sources are centred by construction and carry no centring residual to absorb, and the modes still take the false-detection rate there from 86.17\% to 4.34\%.

The realizations, the summaries and the mode coordinates are released in the Point\_host\_trials, Point\_host\_summary, Point\_only\_rates, Point\_host\_recovery and Mode\_coordinates worksheets of the Source Data.

\subsection{Catalogue-level estimate of the corrected F444W fraction}\label{sec:catalogue-estimate}

What the correction implies for the numbers Zhang et al. publish is estimated here from the object-level catalogue they released, with no new image fitting. Their released residual stack cannot supply an exact corrected measurement, because every point source has already been subtracted from it with its own fitted amplitude, so the freedom that carries the effect is no longer present in it. An exact refit would instead need the local empirical PSF, the mask, the uncertainty map, the neighbour models, the background and the fitted point parameters of each of the 217 objects. Products of that kind are not customarily archived with an imaging paper, and there was no reason to archive them here, since the quantity that requires them is defined by the present work. What follows is therefore a catalogue-level estimate under an explicit model, with every image-level quantity held at the value Zhang et al. published. It proceeds in five steps: (1) each of their 217 objects is given its own nuclear colour, computed from the fluxes in their catalogue; (2) a leakage relation, which converts that colour into the fraction of nuclear light the conventional fit misassigns to a galaxy, is calibrated on the noiseless allocations of the \hyperref[sec:robustness]{robustness test}; (3) the two host measurements Zhang et al. report, one summed over objects and one from their stack, are brought onto a common scale; (4) the misassigned light predicted for each object is removed from its published host flux and what remains is summed; and (5) that sum is converted into a magnitude and a stellar mass.

The nuclear flux of an object in each band is the total flux Zhang et al. publish for it minus the host flux they publish for it. The ratio of its nuclear fluxes in F277W and F444W then provides a two-band proxy for the colour of that nucleus,
\begin{equation}
\beta_j = \frac{\ln\left(p_{j,\mathrm{F444W}}/p_{j,\mathrm{F277W}}\right)}{\ln\left(4.401/2.786\right)}.
\label{eq:object_slope}
\end{equation}
This needs a positive nuclear flux in those two bands only, and 216 of the 217 objects have one. Of the four objects excluded when the representative nuclear spectrum of equation~\eqref{eq:median_flux} was formed, which required positive nuclear flux in all four bands, only one fails in F277W, and that object was kept here and given the median colour of the other 216 so that the sample size is not quietly reduced.

Two properties of that proxy limit what the estimate can claim. A two-band colour does not determine the detailed spectrum inside F444W, so the calculation assumes that the smooth-continuum leakage relation of equation~\eqref{eq:leakage_relation} describes each object adequately, and strong emission lines or spectral breaks are part of its systematic uncertainty. And the nuclear fluxes themselves come from the decomposition being corrected, so they are provisional. Since the leakage grows with wavelength and vanishes in F115W, an over-allocated F444W host lowers the inferred nuclear F444W flux, which flattens the colour and reduces the leakage, so that circularity makes the resulting correction conservative.

The leakage relation is calibrated on the three smooth-continuum cases of the robustness test, in which a synthetic nucleus containing no galaxy at all was fitted with the stellar template and surrendered 15.67\%, 18.78\% and 22.56\% of its light to the Sérsic component at nuclear colours of 3.0, 3.896 and 5.0. A straight line through those three points is anchored at zero leakage for a nucleus whose colour equals that of the stellar reference, $\beta_{\mathrm{star}} = -1.538$. The anchor imposes the one value the relation is known to take exactly, since a source whose spectrum matches its template surrenders nothing. Fitting the line under that anchor gives
\begin{equation}
\ell\left(\beta\right) = \max\left[0,\ 0.0345226\,\beta + 0.0531050\right],
\label{eq:leakage_relation}
\end{equation}
which returns the leaked fraction for any nuclear colour between the stellar value and 5.0. Colours were clipped to that interval before the relation was used, so no object is corrected on an extrapolation beyond the range in which the relation was measured.

The object-level sum and the residual-stack measurement are different estimators, and they give aggregate F444W host-to-total fractions of 13.48\% and 11.8\% respectively. Since the correction is applied object by object but revises the stacked figure, every object-level host flux was first multiplied by 0.8754 to put the two on one scale. The factor is independent of colour and therefore leaves the ranking on which the correction acts unchanged, but the rescaling assumes that the difference between the two estimators is itself independent of source colour and brightness, and that assumption cannot be tested without the individual images.

The light predicted to be misassigned in one object is its leaked fraction at its own colour, multiplied by the nuclear F444W flux Zhang et al. publish for it. That amount is removed from the object's rescaled host flux, and where the removal would leave a negative host it is set to zero, because a galaxy cannot have negative flux. Adding what survives across the 217 objects and dividing by the summed total F444W flux of the sample gives 4.64\%, against the 11.8\% Zhang et al. report. The 95\% interval on that figure runs from 3.23\% to 6.25\% and comes from rebuilding the sample 10,000 times by drawing whole objects with replacement, which keeps each object's colour, nuclear flux and host flux together, as resampling those quantities separately would not. That interval measures sampling variation alone and carries no uncertainty from the two-band colour proxy, from the leakage relation or from the common rescaling factor. The same leakage charged to each object's total flux gives a more aggressive sensitivity estimate of 3.04\%. The reported value is 4.64\%, because the mismatch originates in the nucleus, and the alternative normalization tests how much the answer depends on that choice.

The direction of the correction is safer than its size. For it to point the wrong way, LRD nuclei would have to weight F444W as a stellar spectrum does, and the photometry Zhang et al. publish places their nuclear colours far from the stellar value at which equation~\eqref{eq:leakage_relation} returns nothing.

The catalogue model attributes 7.2 percentage points, which is 61\% of the published F444W component, to the mismatched nuclear template. The 4.64\% that remains makes the inferred host 2.54 times fainter and 1.01 magnitudes dimmer in the rest-frame optical. At a fixed mass-to-light ratio the published mass becomes $\log_{10}(M/M_\odot) = 9.02 + \log_{10}(4.64/11.8) = 8.61$, a reduction of 0.41 dex, and that figure is a rescaling of their value and not an independent population fit. The correction also changes the inferred host colour, because it removes progressively more light towards F444W and none at all in F115W. The resulting spectrum is bluer and would generally favour a younger or less obscured population with a lower mass-to-light ratio, although the age and extinction degeneracy of a four-band fit keeps $10^{8.61}\,M_\odot$ from being a formal upper limit. Independent spectroscopy points the same way. Sun, Naidu and colleagues obtain a host stellar mass of $10^{8.3}\,M_\odot$ from their median stack of 98 little red dots\cite{sun2026}, and the chromatic correction alone moves the imaging estimate from $10^{9.02}$ to $10^{8.61}\,M_\odot$, closing more than half of that logarithmic difference. The two samples and their redshift distributions are not identical, so that agreement is suggestive and it is not a direct validation.

What survives the correction is compatible with a genuine host and does not on its own establish that every LRD contains one, because 4.64\% is a catalogue-level residual and not an image-level refit. The correction changes the inferred luminosity and stellar mass while leaving the host morphology to be settled elsewhere. It also leaves the ratio of black-hole mass to host mass undetermined, because spectroscopic modelling is revising the black-hole masses at the same time and by about a factor of one hundred\cite{rusakov2026}.

The condition applies well beyond this sample. Its coefficients belong to F444W, to the empirical template released with that paper and to the Sérsic bounds those authors imposed, but the three ingredients the calculation needs, a measured nuclear colour, a leakage relation calibrated on hostless simulations and a per-object nuclear flux, exist wherever compact nuclei are decomposed in broad filters. The sign of the effect follows from which way the source and the calibration stars weight the filter. For the red LRD spectra measured here it projects positively onto the allowed Sérsic family in F150W, F277W and F444W, where it can create an apparent host or enlarge a real one, and negatively in F115W, where the non-negative amplitude clamps the allocation to zero and a genuine host can instead be under-recovered. Host measurements around compact sources are therefore conditional on the spectral compatibility of the nucleus with the stars used to build its template, in both directions, and the configuration in which the effect adds light is the long-wavelength one in which rest-frame optical hosts are measured. That configuration is widely used. Zhuang and Shen build their NIRCam templates from field stars and report that a template narrower than the source inflates and compacts the recovered host\cite{zhuang2024}, Chen and colleagues construct an empirical PSF from stacked field stars to search for LRD hosts\cite{chen2025}, and Whalen and colleagues examine the limits of morphological fitting for the same objects with a fixed empirical template\cite{whalen2026}. None of that work conditions the nuclear template on the source spectrum.

One qualitative consistency check is available in the published parameters. Fitting the synthetic source that contains no galaxy drives the Sérsic index to the lower bound of the allowed interval in the two long bands, together with the lower bound in effective radius, and leaves it above that bound in F150W. Table 1 of Zhang et al. shows the same wavelength dependence, with the index at 1.00 $\pm$ 0.14 in F277W and 1.00 $\pm$ 0.12 in F444W and at 2.00 $\pm$ 1.04 and 1.66 $\pm$ 0.70 in the two shorter bands. Those fitted indices played no part in constructing the correction, so the shared boundary behaviour is a consistency check, and it is a behaviour that other causes could also produce.

The per-object corrections, the published individual ratios, the aggregate summary and the itemised inventory of the products an exact refit would require are in the Catalogue\_correction, Published\_individual\_ratios, Catalogue\_correction\_summary and Refit\_requirements worksheets of the Source Data.

\section*{Use of AI-based language tools}

The author used AI-based language tools to assist with English translation and stylistic revision of the manuscript. All scientific content, interpretations, calculations and conclusions are solely the author's own.

\section*{Data availability}

Source Data are provided with this paper and are also deposited at Zenodo under DOI 10.5281/zenodo.22182425. The article and numerical Source Data released by Zhang et al. are available with ref.~\citenum{zhang2026}. The COSMOS-Web imaging is available from the survey release at https://cosmos2025.iap.fr/ and is described in ref.~\citenum{casey2023}. Castelli and Kurucz ATLAS9 model atmospheres are available from the public grid described in ref.~\citenum{castelli2004}. The accompanying data package contains the values reported in Table~\ref{tab1} and Extended Data Table~\ref{tabED1}, every numerical series plotted in the figures, the complete 100-pair injection table, the controlled point-and-host realizations, the matched host-preservation estimands, the model-atmosphere slopes, the 324-point Sérsic-centroid grid and the 217-row catalogue correction, with a README worksheet mapping every table to the analysis that produced it. Original third-party images and model grids remain in their cited repositories and are identified by source.

\section*{Code availability}

The analysis used Python 3.14.5 with NumPy 2.3.5, SciPy, Astropy and synphot, together with STPSF 2.2.0\cite{perrin2014,rieke2023}, galight 0.2.1\cite{ding2021} and lenstronomy 1.11.10\cite{birrer2021}, each publicly available from its own distribution under its own licence. Every operation applied to them is stated as an equation in the Methods.

\section*{Author contributions}

S.B.-G. conceived the study, developed the methodology and software, performed the analysis, interpreted the results, prepared the figures and wrote the manuscript.

\section*{Competing interests}

The author declares no competing interests.

\section*{Acknowledgements}\label{Acknowledgements}

Sergio Bonaque-Gonz{\'a}lez holds a postdoctoral contract ``Viera y Clavijo'' funded by the Agencia Canaria de Investigaci{\'o}n, Innovaci{\'o}n y Sociedad de la Informaci{\'o}n.

\bibliography{sn-bibliography}

\section*{Tables}

\begin{table}[htbp]
\caption{Sérsic allocations measured for a source containing no galaxy}\label{tab1}
\begin{tabular}{@{}lllll@{}}
\toprule
Filter & Published extended fraction & Hostless-source allocation & Ratio to published & Redistributed P-PSF light \\
\midrule
F115W & 39.6\% & 0.00\% & 0.00 & 0.02\% \\
F150W & 33.9\% & 2.62\% & 0.08 & 0.39\% \\
F277W & 23.2\% & 6.58\% & 0.28 & 0.81\% \\
F444W & 11.8\% & 18.73\% & 1.59 & 1.72\% \\
\botrule
\end{tabular}
\footnotetext{The published fraction is measured by Zhang et al. on the stacked residual in each filter. The hostless-source allocation is the minimum-$\chi^2$ Sérsic fraction obtained when a synthetic unresolved LRD P-PSF containing zero host flux is fitted with the empirical stellar P-PSF and the bounded Sérsic family. The ratio divides the hostless allocation by the published fraction. Redistributed P-PSF light is one half of the summed absolute pixel difference between the normalized LRD and stellar P-PSFs. No random noise is added, so no sampling uncertainty is assigned to the hostless-source allocation.}
\end{table}

\begin{table}[htbp]
\renewcommand{\tablename}{Extended Data Table}
\renewcommand{\thetable}{1}
\caption{Recovered galaxy fraction for nine injected truths}\label{tabED1}
\begin{tabular}{@{}lllllll@{}}
\toprule
\begin{tabular}[t]{@{}l@{}}Injected\\fraction (\%)\end{tabular} &
\begin{tabular}[t]{@{}l@{}}Injected $R_{\mathrm{e}}$\\(arcsec)\end{tabular} &
\begin{tabular}[t]{@{}l@{}}Matched\\control (\%)\end{tabular} &
\begin{tabular}[t]{@{}l@{}}Conventional\\fit (\%)\end{tabular} &
\begin{tabular}[t]{@{}l@{}}Source-\\conditioned (\%)\end{tabular} &
\begin{tabular}[t]{@{}l@{}}Excess over\\control\end{tabular} &
\begin{tabular}[t]{@{}l@{}}Residual after\\correction\end{tabular} \\
\midrule
5  & 0.030  & 3.8  & 18.3 & 3.4  & \begin{tabular}[t]{@{}l@{}}$+14.79$\\$[10.69, 19.28]$\end{tabular} & \begin{tabular}[t]{@{}l@{}}$-0.10$\\$[-0.17, 0.02]$\end{tabular} \\
5  & 0.0375 & 4.1  & 15.4 & 4.0  & \begin{tabular}[t]{@{}l@{}}$+11.95$\\$[8.90, 13.98]$\end{tabular}  & \begin{tabular}[t]{@{}l@{}}$-0.05$\\$[-0.20, 0.04]$\end{tabular} \\
5  & 0.050  & 4.5  & 13.0 & 4.4  & \begin{tabular}[t]{@{}l@{}}$+9.45$\\$[7.16, 10.01]$\end{tabular}   & \begin{tabular}[t]{@{}l@{}}$-0.05$\\$[-0.22, 0.04]$\end{tabular} \\
12 & 0.030  & 7.7  & 23.9 & 6.9  & \begin{tabular}[t]{@{}l@{}}$+17.09$\\$[13.65, 17.70]$\end{tabular} & \begin{tabular}[t]{@{}l@{}}$-0.50$\\$[-0.92, -0.25]$\end{tabular} \\
12 & 0.0375 & 8.6  & 21.5 & 8.4  & \begin{tabular}[t]{@{}l@{}}$+12.77$\\$[11.78, 13.08]$\end{tabular} & \begin{tabular}[t]{@{}l@{}}$-0.26$\\$[-0.53, -0.13]$\end{tabular} \\
12 & 0.050  & 10.0 & 19.1 & 9.8  & \begin{tabular}[t]{@{}l@{}}$+9.05$\\$[8.86, 9.27]$\end{tabular}    & \begin{tabular}[t]{@{}l@{}}$-0.15$\\$[-0.27, -0.11]$\end{tabular} \\
25 & 0.030  & 18.9 & 34.5 & 17.2 & \begin{tabular}[t]{@{}l@{}}$+14.86$\\$[14.28, 15.24]$\end{tabular} & \begin{tabular}[t]{@{}l@{}}$-0.78$\\$[-1.51, -0.70]$\end{tabular} \\
25 & 0.0375 & 21.4 & 32.7 & 20.0 & \begin{tabular}[t]{@{}l@{}}$+10.99$\\$[10.80, 11.24]$\end{tabular} & \begin{tabular}[t]{@{}l@{}}$-0.68$\\$[-1.02, -0.28]$\end{tabular} \\
25 & 0.050  & 22.9 & 30.7 & 22.4 & \begin{tabular}[t]{@{}l@{}}$+7.70$\\$[7.60, 8.00]$\end{tabular}    & \begin{tabular}[t]{@{}l@{}}$-0.27$\\$[-0.62, -0.10]$\end{tabular} \\
\botrule
\end{tabular}
\footnotetext{Each row is one injected truth, measured on 32 trials drawn from eight objects with four background patches each. The injected fraction and effective radius define the circular exponential galaxy added around the nucleus. The three recovered columns are medians of the percentage of total fitted flux assigned to the galaxy, for the matched control whose nucleus carries the stellar P-PSF, for the conventional fit in which a chromatic nucleus is fitted with the stellar template, and for the source-conditioned fit in which that same nucleus is fitted with its own P-PSF. The last two columns are medians of the per-trial difference from the matched control in percentage points, with 95\% object-cluster bootstrap intervals beneath them. Because these differences are calculated for each paired trial before taking the median, they need not equal the differences between the displayed column medians. The matched control lies below the injected fraction in every row because subtracting the background removes some genuine galaxy light as well, so the correction is judged against the control and not against the injected value.}
\end{table}

\typeout{get arXiv to do 4 passes: Label(s) may have changed. Rerun}
\end{document}